\documentclass[journal]{IEEEtran}

\usepackage{booktabs}
\usepackage{xcolor}
\usepackage{colortbl}
\usepackage{amssymb}
\usepackage{algorithmic}
\usepackage{amsmath} 
\usepackage{amsfonts}

\ifCLASSINFOpdf
\usepackage[pdftex]{graphicx}
\else
\usepackage[dvips]{graphicx}
\fi
\usepackage{epstopdf}
\usepackage{array}
\ifCLASSOPTIONcompsoc
\usepackage[caption=false,font=normalsize,labelfont=sf,textfont=sf]{subfig}
\else
\usepackage[caption=false,font=footnotesize]{subfig}
\fi
\usepackage{float} 
\usepackage{fixltx2e}
\usepackage{color}
\usepackage{xcolor}
\usepackage{textcomp}
\usepackage[normalem]{ulem}
\usepackage{cite}
\usepackage{multibib} 
\usepackage{nomencl}
\makenomenclature
\usepackage{etoolbox}
\renewcommand\nomgroup[1]{%
	\item[\bfseries
	\ifstrequal{#1}{A}{Indices and Sets}{%
		\ifstrequal{#1}{B}{Continuous Decision Variables}{%
			\ifstrequal{#1}{C}{Discrete Decision Variables}{       \ifstrequal{#1}{D}{Parameters}{
			        \ifstrequal{#1}{E}{Dynamic System Variables and Parameters}} }}} ]}

\usepackage{hyperref}
\hypersetup{
	colorlinks=true,
	linkcolor=blue,
	filecolor=magenta,      
	urlcolor=cyan,}
\IEEEoverridecommandlockouts

\begin{document}
\bstctlcite{BSTcontrol}
	
\title{Approximating Trajectory Constraints with Machine Learning — Microgrid Islanding with Frequency Constraints}

\author{Yichen~Zhang,~\IEEEmembership{Member,~IEEE,}
	~Chen~Chen,~\IEEEmembership{Senior Member,~IEEE,}
	~Guodong~Liu,~\IEEEmembership{Senior Member,~IEEE,}
	~Tianqi~Hong,~\IEEEmembership{Member,~IEEE,}
	~Feng~Qiu,~\IEEEmembership{Senior Member,~IEEE}
	\thanks{
	This work was supported by the U.S. Department of Energy Office of Electricity -- Advanced Grid Modeling Program.
	
	Y. Zhang, T. Hong, F. Qiu are with Argonne National Laboratory, Lemont, IL 60439 USA (email: \{yichen.zhang, thong, fqiu\}@anl.gov).
	
	C. Chen was with Argonne National Laboratory, Lemont, IL 60439 USA (email: morningchen@anl.gov).
	
	G. Liu is with Oak Ridge National Laboratory, Oak Ridge, TN 37831 USA (email: liug@ornl.gov).
}}

\markboth{This paper has been accepted by IEEE TRANSACTIONS ON POWER SYSTEMS (DOI: 10.1109/TPWRS.2020.3015913)}%
{Shell \MakeLowercase{\textit{et al.}}: Bare Demo of IEEEtran.cls for IEEE Journals}
\maketitle

\begin{abstract}
In this paper, we introduce a deep learning aided constraint encoding method to tackle the frequency-constraint microgrid scheduling problem. The nonlinear function between system operating condition and frequency nadir is approximated by using a neural network, which admits an exact mixed-integer formulation (MIP). This formulation is then integrated with the scheduling problem to encode the frequency constraint. With the stronger representation power of the neural network, the resulting commands can ensure adequate frequency response in a realistic setting in addition to islanding success. The proposed method is validated on a modified 33-node system. Successful islanding with a secure response is simulated under the scheduled commands using a detailed three-phase model in Simulink. The advantages of our model are particularly remarkable when the inertia emulation functions from wind turbine generators are considered.
\end{abstract}

%

\mbox{}
\nomenclature[A,00]{$h$, $\mathcal{N}_{\text{P}}$, $N_{\text{P}}$}{index, set, number of substations}
\nomenclature[A,01]{$i$, $\mathcal{N}_{\text{D}}$, $N_{\text{D}}$}{index, set, number of diesel generators}
\nomenclature[A,02]{$j$, $\mathcal{N}_{\text{W}}$, $N_{\text{W}}$}{index, set, number of wind turbine generators}
\nomenclature[A,03]{$k$, $\mathcal{N}_{\text{B}}$, $N_{\text{B}}$}{index, set, number index of buses}
\nomenclature[A,05]{$l$, $\mathcal{N}_{\text{L}}$, $N_{\text{L}}$}{index, set, number index of lines}
\nomenclature[A,06]{$m$, $\mathcal{N}_{\text{Y}}$, $N_{\text{Y}}$}{index, set, number of hidden layers of a neural network}
\nomenclature[A,07]{$n$, $\mathcal{N}_{\text{O}}$, $N_{\text{O}}$}{index, set, number of neurons in a layer}
\nomenclature[A,09]{$s$, $\mathcal{N}_{\text{S}}$, $N_{\text{S}}$}{index, set, number of training samples}
\nomenclature[A,10]{$t$, $\mathcal{N}_{\text{T}}$, $N_{\text{T}}$}{index, set, number of periods}

\nomenclature[B,01]{$P_{i,t}^{\text{D}}$}{power output of diesel generator $i$ from its minimum during period $t$}
\nomenclature[B,02]{$p_{i,t}^{\text{D}}$}{incremental output of diesel generator $i$ from its minimum during period $t$}
\nomenclature[B,03]{$R_{i,t}^{\text{D}}$}{reserve of diesel generator $i$ during period $t$}
\nomenclature[B,04]{$P_{l,t}$, $Q_{l,t}$}{active, reactive power flow on line $l$ during period $t$}
\nomenclature[B,05]{$P_{h,t}^{\text{PCC}}$}{power flow at point of common coupling $h$ during period $t$}
\nomenclature[B,06]{$V_{k,t}$}{voltage of bus $k$ during period $t$}

\nomenclature[C,01]{$u_{i,t}^{\text{D}}$, $u_{j,t}^{\text{W}}$}{1 if unit $i$, $j$ is scheduled on during period $t$ and 0 otherwise}
\nomenclature[C,02]{$u_{j,t}^{\text{IE}}$}{1 if inertia emulation of unit $j$ is scheduled on during period $t$ and 0 otherwise}

\nomenclature[D,01]{$P_{k,t}^{\text{L}}$, $Q_{k,t}^{\text{L}}$}{active, reactive power demand at bus $k$ during period $t$}
\nomenclature[D,02]{$P_{j,t}^{\text{W}}$}{power output of wind turbine generator $j$ during period $t$}
\nomenclature[D,02]{$\overline{P}_{j}^{\text{W}}$}{rated power output of wind turbine generator $j$ at fully loaded condition}
\nomenclature[D,03]{$\underline{P}_{l}$, $\overline{P}_{l}$}{min, max active power flow of line $l$}
\nomenclature[D,04]{$\underline{Q}_{l}$, $\overline{Q}_{l}$}{min, max reactive power flow of line $l$}
\nomenclature[D,05]{$\underline{P}^{\text{D}}_{i}$, $\overline{P}^{\text{D}}_{i}$}{min, max active power output of unit $i$}
\nomenclature[D,06]{$\underline{Q}^{\text{D}}_{i}$, $\overline{Q}^{\text{D}}_{i}$}{min, max reactive power output of unit $i$}
\nomenclature[D,07]{$\bold{W}_{m}$, $\bold{b}_{m}$}{weight and bias of layer $m$ in a neural network}
\nomenclature[D,08]{$\rho^{\text{PCC}}_{t}$}{purchasing price of energy from distribution grid during period $t$}
\nomenclature[D,09]{$\lambda^{\text{F}}_{i}$}{fixed cost of unit $i$ at the point of $\underline{P}^{\text{G}}_{i}$}
\nomenclature[D,10]{$\lambda^{\text{M}}_{i}$}{marginal cost of unit $i$}
\nomenclature[D,11]{$\lambda^{\text{S}}_{i}$}{start-up cost of unit $i$}
\nomenclature[D,12]{$\epsilon$}{allowable voltage deviation from nominal value}

\nomenclature[E,02]{$\psi_{ds}$,$\psi_{qs}$}{stator flux linkage in $d$, $q$-axis}
\nomenclature[E,02]{$\psi_{dr}$,$\psi_{qr}$}{rotor flux linkage in $d$, $q$-axis}
\nomenclature[E,03]{$v_{ds}$, $v_{qs}$}{instantaneous stator voltage in $d$, $q$-axis}
\nomenclature[E,03]{$v_{dr}$, $v_{qr}$}{instantaneous rotor voltage in $d$, $q$-axis}
\nomenclature[E,03]{$i_{ds}$, $i_{qs}$}{instantaneous stator current in $d$, $q$-axis}
\nomenclature[E,03]{$i_{dr}$, $i_{qr}$}{instantaneous rotor current in $d$, $q$-axis}
\nomenclature[E,04]{$L_{m}$}{mutual inductance}
\nomenclature[E,04]{$R_{s}$, $L_{ls}$}{stator resistance, leakage inductance}
\nomenclature[E,04]{$R_{r}$, $L_{lr}$}{rotor resistance, leakage inductance}
\nomenclature[E,05]{$\overrightarrow{\Psi_{s}}$, $\Psi_{s}$}{space vector of stator flux and its magnitude}
\nomenclature[E,06]{$\overrightarrow{V_{s}}$, $V_{s}$}{space vector of stator voltage and its magnitude}
\nomenclature[E,07]{$H_{D}$, $H_{T}$}{diesel, wind turbine  inertia constant(s)}
\nomenclature[E,08]{$T_{m}$}{mechanical torque of wind turbine generators}
\nomenclature[E,08]{$P_{m}$, $P_{e}$}{mechanical, electric power of diesel generators}
\nomenclature[E,08]{$P_{g}$, $Q_{g}$}{active, reactive power of wind turbines}
\nomenclature[E,08]{$P_{v}$}{valve position of diesel generators}
\nomenclature[E,08]{$R_{D}$}{governor droop setting of diesel generators}
\nomenclature[E,09]{$\tau_{d}$, $\tau_{sm}$}{diesel engine, governor time constant(s)}
\nomenclature[E,11]{$\omega_{d}$, $\omega_{r}$}{diesel, wind turbine angular speed}
\nomenclature[E,13]{$\omega_{s}$}{synchronous angular speed}
\nomenclature[E,14]{$\overline{\omega}$}{speed base of wind turbine generator (rad/s)}
\nomenclature[E,15]{$\overline{f}$}{speed base of diesel generator (Hz)}
\printnomenclature[0.7in]

\section{Introduction}\label{sec_intro}
Microgrids have proven to be a versatile way to improve grid resiliency\cite{Chen2016}\cite{Wang2015a}. When the main grid undergoes extreme events\cite{Ju2017}, transitioning into islanded operation using microgrids enables uninterrupted and sustained customer supply. An islanding demand can be issued in different time scales — from seconds to minutes to tens of minutes — and islanding events can be generally categorized into scheduled and event-triggered. Scheduled islanding is issued in a slower time scale, that is, minutes to tens of minutes, and is generally for economic and maintenance purposes. In these cases, the power at point of commend coupling (PCC) will be controlled to a sufficiently small value before opening the breakers. On the other hand, event-triggered islanding is usually executed by local logic to automatically isolate microgrids from faults at the main grid. The islanding must be executed immediately upon the occurrence of faults, and thus is in the time scale of seconds \cite{Xu2012}. It is event-triggered islanding that enhances  grid resiliency and is thus our focus. 

In event-triggered islanding, the absent PCC power will result in deviation of frequency and voltage trajectories. Regaining power balance while maintaining trajectory deviation within permissible ranges is the key to successful islanding. Unsuccessful islanding may occur due to inadequate capacity for power sharing, loss of synchronization for grid-interactive inverters, and/or small-signal instability. Given these factors, preventive scheduling of microgrids is essential. Islanding capability under load and renewable uncertainty is studied in \cite{Liu2017} and \cite{Gholami2019}. The key concern in these works is to have adequate spinning reserve to ensure proper load sharing (or droop response). Load shedding strategies have also been identified as effective approaches to the microgrid islanding transition. A robust strategy is proposed in \cite{Liu2016}, and an intelligent load shedding approach is studied in \cite{Balaguer2011}, where the optimal amount of load to be shed is computed. While effective, load shedding schemes cause interruptions of customer supply and are thus less desirable \cite{yichen_RAS}. Ref. \cite{Pan2019} probes small-signal stability  with respect to the system operating condition. Frequency response of droop-controlled inverters at steady state after islanding is considered in \cite{Amirioun2018b}, where the droop control gains are co-optimized with other control commands. Ref. \cite{Gholami2018} makes a further improvement by considering the dynamic frequency response constraint in addition to the steady state. In this paper, we also tackle the scheduling problem subject to islanding capability and dynamic frequency response constraints.

Frequency-constrained scheduling in transmission systems has been extensively studied  \cite{Restrepo2005,Lee2013,Chavez2014,Ahmadi2014,Wen2016}. Most studies employ a low-order frequency response model \cite{Anderson1990} to represent system frequency response characteristics, from which an analytical expression of frequency nadir under a step input can be derived \cite{Ahmadi2014}. This expression, which is highly nonlinear, maps the system states and control actions to the frequency nadir.   Piece-wise linearization is then applied to encode this nadir expression into the optimization model. This method allows a tractable computation model, which is shown to be accurate in synchronous generator-dominated bulk grids \cite{Shi2018}. One of the disadvantages of this approach is that it is unable to incorporate responses from grid-interactive inverters, since there is no analytical expression for the step response of higher-order differential equations. But grid-interactive inverters contribute a considerable percentage of frequency response and cannot be omitted. Other practical factors, such as phase-lock loop (PLL) transient, low-pass filters, dead-band, and saturation, will also alter the response and have not been considered yet. 

Motivated by these issues, we introduce here a deep learning aided constraint encoding approach based on \cite{Say2017}. The key is to first parameterize the complex map from system states and controls to the frequency nadir, using the neural network, and then formulate the trained neural network into a mixed-integer linear program. If the rectified linear unit (ReLU) is employed as the activation function, this reformulation is exact \cite{Anderson2019}. Both simulation data and real operational measurements can be used to train the model. A similar idea has been employed in other power system applications, such as voltage control \cite{Hong2019a}. In this paper, we present our approach in the following steps:
\begin{enumerate} 
	\item From the original three-phase network (TPN) model, we extract a positive-sequence power balance (PSPB) model to efficiently generate training data. Dynamic simulations show that the frequency response characteristics of these two models correspond closely to each other.
	\item We introduce the deep learning aided constraint encoding approach to handle the complicated dynamic-constrained optimization problem and discover sufficient features to admit an accurate representation. 
	\item We integrate the deep neural network with the microgrid scheduling model as a mixed-integer linear program (MILP) to perform frequency-constrained energy management where detailed wind turbine generator characteristics are considered. The effectiveness of the constraint encoding technique is verified by detailed TPN simulation.
\end{enumerate}
The advantages of our proposed approach are twofold. On the one hand, various distributed energy resources with different grid-supportive functions can be considered, and model orders and nonlinearity can be incorporated. On the other hand, the neural network model with ReLU activation function admits an exact reformulation compared with other nonlinear regression models. Therefore, our approach is not subject to approximation error when being incorporated into the optimization model.
The remainder of the paper is organized as follows: Section \ref{sec_freq} discusses the frequency response in microgrids and derives the positive-sequence power balance model. Section \ref{sec_dnn} details the training model and MILP encoding. Section \ref{sec_problem} presents the scheduling problem formulation. In Section \ref{sec_case}, training and scheduling results with dynamic simulation verification, including model validation, are presented.

\section{Microgrid Frequency Response in Islanding Events}\label{sec_freq}
\begin{figure}[h]
	\centering
	\includegraphics[scale=0.28]{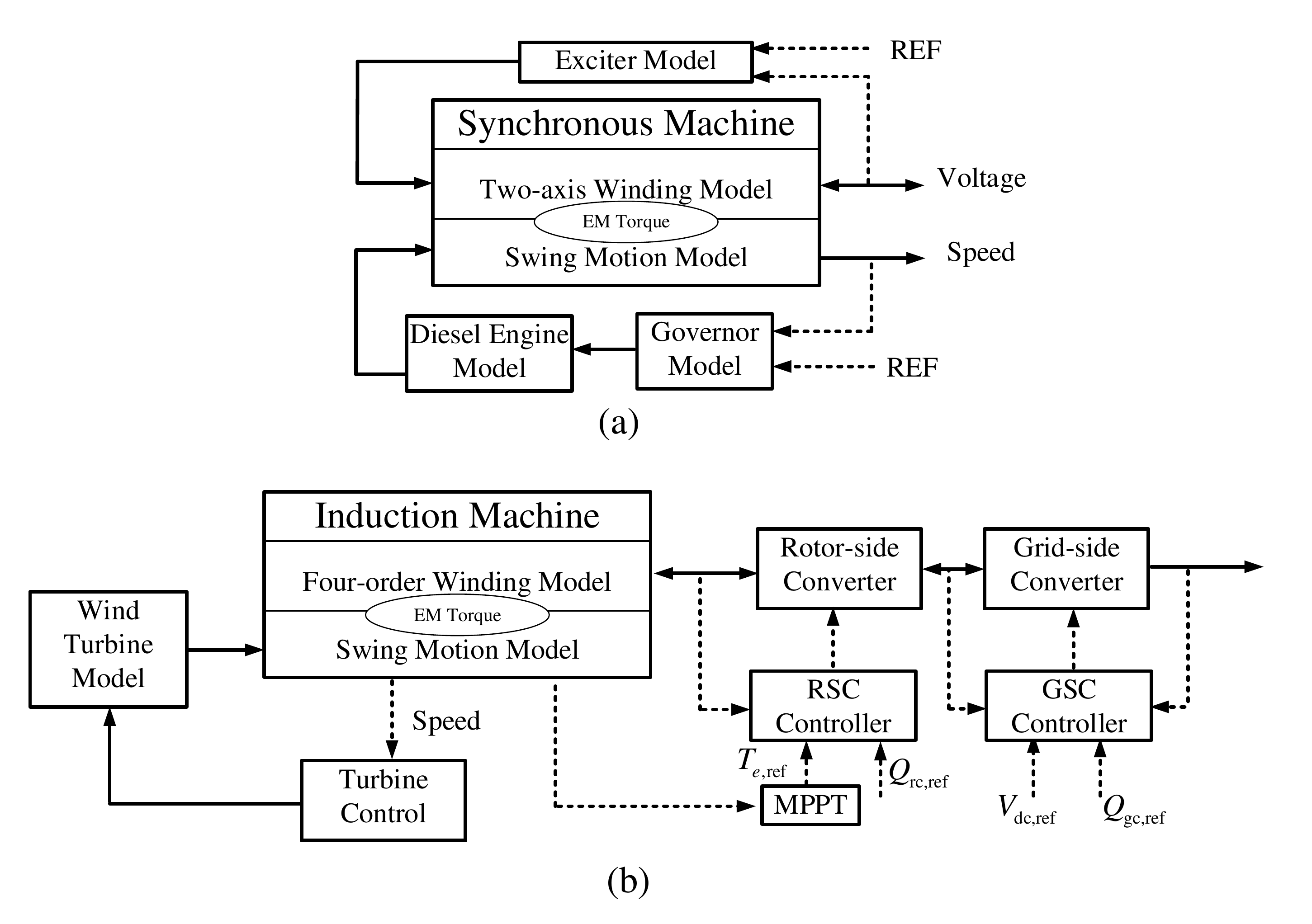}
	\caption{Modules considered in DSGs and WTGs.}
	\label{fig_module}
\end{figure}

In this section, we identify components that contribute to the frequency response in an islanded microgrid,  and we extract a simplified positive-sequence model to sufficiently represent the original system such that the simulation time can be significantly reduced. The microgrid studied in this paper consists of diesel generators (DSGs) and wind turbine generators (WTGs).  Among all distributed energy resources (DERs), WTGs have the most complicated dynamics, so we chose to include them so that our study applies to the most complex microgrids. The TPN model employs three-phase power flow and  positive-sequence machine models. Standard synchronous generator components are also included:  turbine, governor, exciter, and a fifth-order synchronous generator. The WTG uses a double-fed induction machine (DFIG) based configuration and is controlled by a standard field-oriented approach.  The detailed modules considered in DSGs and WTGs are shown in Fig. \ref{fig_module}. The PSPB model consists of power variation models for the DSG and WTG, for disturbances and controls, and a power balance constraint.

We assume that at least one diesel generator operates in both grid-tied and islanded modes to seamlessly establish frequency and voltage for the islanded microgrid. As DSGs are the only grid-forming source in the islanded microgrid, built-in speed governors dominate the frequency response in an islanded microgrid, as described by the following equations:
\begin{align}
\label{eq_DSG}
\begin{aligned}
2H_{D}\Delta\dot{\omega}_{d}&=\overline{f}(\Delta P_{m}- \Delta P_{e})\\
\tau_{d}\Delta\dot{P}_{m}&=-\Delta P_{m}+\Delta P_{v}\\
\tau_{sm}\Delta\dot{P}_{v}&= -\Delta P_{v}  - \Delta\omega_{d}/(\overline{f}R_{D})
\end{aligned}
\end{align}
Eq. (\ref{eq_DSG}) can represent the frequency response of either a single DSG or an aggregated group. When representing an aggregation, this model represents the averaged frequency trajectory of all machines across the entire network. The frequency trajectory of each machine will deviate from the average trajectory to a certain extent. The trajectory distance is determined by the electric distances between different machines and their parameter variations. Ref. \cite{Anderson1990} and \cite{Zhang2018a} have shown that such an aggregation is accurate; that is, the trajectory deviation is sufficiently small in small networks and microgrids, due to the closer electric distance. In the aggregation form, $H_{D}$ is the center of inertia, which can be calculated as follows:
\begin{align}\label{eq_COI}
\begin{aligned}
H_{D}=\dfrac{\sum_{i=1}^{N_{D}}S_{i}^{s}H_{i}^{s}}{S_{\text{sg}}},S_{\text{sg}}=\sum_{i\in \mathcal{S}}^{N_{s}}S_{i}^{s}
\end{aligned}
\end{align}
where $S_{i}^{s}$ and $H_{i}^{s}$ are the base and inertia constant of DSG $i$, respectively. Note that the total power base is the sum of the base of each DSG, and therefore changes according to the commitment command. The detailed procedures to average other parameters, such as time constants of turbines and governors, can be found in \cite{Egido2009,Apostolopoulou2016,Shi2018}. The remaining parts of a DSG, such as the exciter and flux linkages dynamics, though important in TPN simulation, have negligible impacts on frequency response.
\begin{figure*}[h]
	\centering
	\includegraphics[scale=0.3]{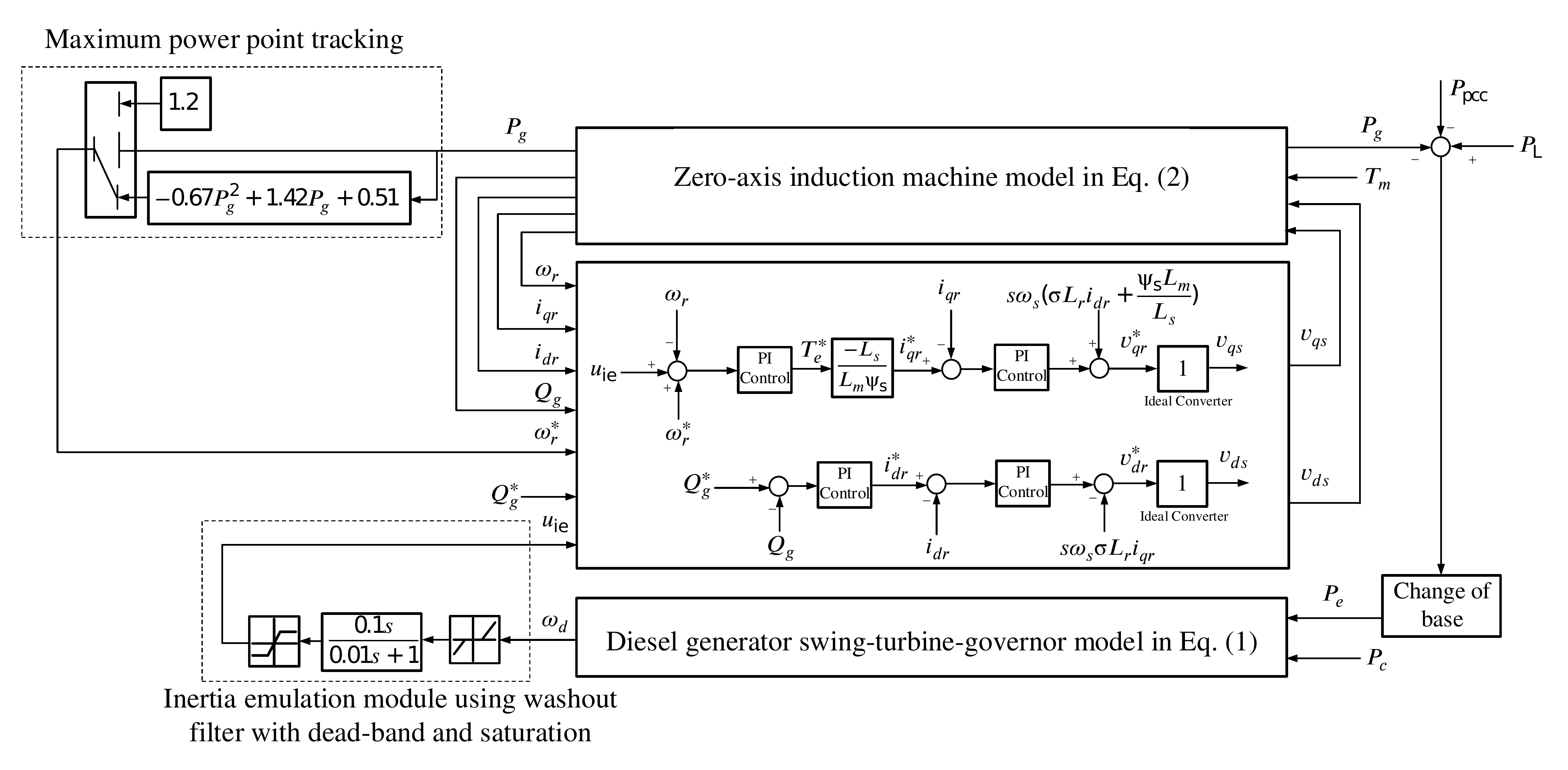}
	\caption{The positive-sequence power balance (PSPB) model for generating training data.}
	\label{fig_PSPB}
\end{figure*}
WTGs are assumed to provide inertia emulation functions. A double-fed induction machine (DFIG)-based WTG consists of a wind turbine, induction machine, rotor-side converter (RSC) and grid-side converter (GSC).  The computation of mechanical torque can be found in \cite{hector}.

The dynamics most relevant to inertia emulation are the induction machine and its speed regulator via the RSC, which are in the electro-mechanical time scale. The control of DC-link voltage and GSC are close to electromagnetic \cite{Tang2018}, and thus omitted. We employ the symmetrical induction machine in the $dq$ reference frame, and the dynamics of the flux linkages are assumed to be infinitely fast, which renders to algebraic constraints. This is called the zero-axis machine model \cite{hector,sauer_chow_book_2017}. The model just described is expressed as follows:
\begin{align}\label{eq_WTG}
\begin{aligned}
&\dot{\omega}_{r}=1/(2H_{T})[T_{m}-\frac{L_{m}}{L_{s}}(\psi_{qs}i_{dr}-\psi_{ds}i_{qr})]\\
&0= \overline{\omega}(v_{qs} - R_{s}i_{qs} - \omega_{s}\psi_{ds})\\
&0= \overline{\omega}(v_{ds} - R_{s}i_{ds} + \omega_{s}\psi_{qs})\\
&0= \overline{\omega}[v_{qr} - R_{r}i_{qr} - (\omega_{s}-\omega_{r})\psi_{dr}]\\
&0= \overline{\omega}[v_{dr} - R_{r}i_{dr} + (\omega_{s}-\omega_{r})\psi_{qr}]\\
&0=-\psi_{qs}+L_{s}i_{qs} + L_{m}i_{qr}\\
&0=-\psi_{ds}+L_{s}i_{ds} + L_{m}i_{dr}\\
&0=-\psi_{qr}+L_{r}i_{qr} + L_{m}i_{qs}\\
&0=-\psi_{dr}+L_{r}i_{dr} + L_{m}i_{ds}\\
&0=P_{g}+(v_{qs}i_{qs}+v_{qs}i_{qs}) + (v_{qr}i_{qr}+v_{qr}i_{qr})\\
&0=Q_{g}+(v_{qs}i_{ds}-v_{ds}i_{qs}) + (v_{qr}i_{dr}-v_{dr}i_{qr})\\
\end{aligned}
\end{align}
where $L_{s}=L_{ls}+L_{m}$ and $L_{r}=L_{lr}+L_{m}$. With the two machine models in (\ref{eq_DSG}) and (\ref{eq_WTG}), the overall PSPB model is shown in Fig. \ref{fig_PSPB}.

The RSC control receives the measurement and uses the field-oriented control (FOC) scheme. By aligning the stator flux vector $\overrightarrow{\Psi_{s}}$ with the direct axis ($d$ axis) of the reference frame, the active power can be controlled independently by the rotor-side quadrature current $i_{qr}$ \cite{DFIM2011}. The response time of the converters is considered to be infinitely fast such that the commands equal the outputs; that is, $v_{qr}=v_{qr}^{*}$ and $v_{dr}=v_{dr}^{*}$. The standard MPPT is designed in such a way that if $P_{g}\leq \overline{P}^{\text{W}}$, the optimal speed is equal to $-0.67P_{g}^{2}+1.42P_{g}+0.51$, and otherwise to 1.2 \cite{Zhang2018a}. The inertia emulation function requires the rate of change of frequency (ROCOF) as the input signal. For physical realization, a washout filter $K_{\text{ie}}s/(0.01s+1)$, where $K_{\text{ie}}$ is the inertia emulation gain, is employed to generate this signal. A dead-band is used to prevent the WTG from responding to small frequency fluctuations \cite{Zhang2019}. The dead-band is set to have a sufficiently large upper bound so that the WTGs will not respond to over-frequency events. 

Finally, the power balance is enforced: The power at PCC is set to zero when the simulation reaches steady state to simulate the islanding event. At that point, subtracting wind power output from the total load yields the power output of DSGs. A change of base is made based on the number of on-line DSGs before feeding the power to Eq. (\ref{eq_DSG}). In general, the PSPB captures the core features of the frequency response \cite{Zhang2018a}. For the sake of simplicity without losing generality, only a constant power load model is considered.

\section{Data Driven Constraint Encoding Using Neural Network Models}\label{sec_dnn}
\subsection{Deep Learning Based Frequency Nadir Prediction}\label{sec_sub_dnn_train}
In this paper, we would like to confine the maximum frequency decline during the entire time window of inertial and primary responses. The maximum deviation value is denoted as the frequency nadir. To prevent the scenario in which the nadir is within the permissible range but the system becomes unstable in the following cycles, we will first simulate the model for a time window of ten seconds after the islanding event, and then retrieve the nadir information over the entire simulated trajectory. With this strategy, an unstable frequency trajectory can be identified if an abnormally low nadir value is obtained. Given the load forecast, the scheduling problem will produce a control command $\bold{u}$, which results in a system operation point $\bold{x}$. Under a disturbance $d$, the frequency nadir can be expressed as a nonlinear function of these variables:
\begin{align}
f_{\text{ndr}} = f(\bold{x},\bold{u},d)\label{eq_nonlinear}
\end{align}
The objective is to use a parameterized function, a neural network in this case, to approximate the function in (\ref{eq_nonlinear}). The correct choice of input features among $\bold{u}$, $\bold{x}$ and $d$ is crucial for this task. In our system, when an islanding event occurs, the power at PCC drops to zero, creating a power imbalance to the system. Therefore, the power at PCC $P^{\text{PCC}}_{s}$ under nominal conditions is the disturbance. The dispatch commands related to frequency response are the commitments of DGs $u^{\text{D}}_{i,s}$ and activation of inertia emulation in WTGs $u^{\text{IE}}_{i,s}$. As stated, the frequency characteristic of DSGs is linear. Thus, the actual outputs of DSGs are less relevant as long as adequate reserves are scheduled. The same conclusion can be drawn for the inertia emulation cases, since this function allows activating only in fully loaded conditions. Thus, the on and off status of DSGs and inertia emulation of WTGs are selected as features in addition to the PCC power.

Let $\bold{X}$ denote the input data to the DNN shown in Eq. (\ref{eq_dnn_X_1}): Each row represents a data sample consisting of the PCC power, the on/off status of each DG, and the number of activated inertia emulation functions, denoted by a vector $\bold{x}_{s}=\left[u^{\text{D}}_{1,s},\cdots,u^{\text{D}}_{N_{\text{D}},s},\sum_{j}u^{\text{IE}}_{j,s},P^{\text{PCC}}_{s}\right]$, and each column represents different samples of one variable. Let a vector $\bold{y}$ denote the label output data shown in Eq. (\ref{eq_dnn_y}), which consists of the frequency nadir of different samples.
\begin{align}
\bold{X} &= \left[\begin{array}{c}\bold{x}_{1}^{T},\bold{x}_{2}^{T} ,\cdots  \bold{x}^{T}_{N_{\text{S}}} \end{array}\right]^{T} \label{eq_dnn_X_1}\\
&=\left[\begin{array}{ccccccc}
							u^{\text{D}}_{1,1} & \cdots & u^{\text{D}}_{N_{\text{G}},1} & \sum_{j}u^{\text{IE}}_{j,1} & P^{\text{PCC}}_{1}\\
							u^{\text{D}}_{1,2} & \cdots & u^{\text{D}}_{N_{\text{G}},2} & \sum_{j}u^{\text{IE}}_{j,2} & P^{\text{PCC}}_{2}\\
							\vdots & \ddots & \vdots & \vdots &  \vdots\\
							u^{\text{D}}_{1,N_{\text{S}}} & \cdots & u^{D}_{N_{\text{G}},N_{\text{S}}} & \sum_{j}u^{\text{IE}}_{j,N_{\text{S}}} & P^{\text{PCC}}_{N_{\text{S}}}\\
							\end{array}\right] \label{eq_dnn_X_2}
\end{align}
\begin{align}
\label{eq_dnn_y}
\bold{y}^{*}= \left[\begin{array}{ccccc}
f_{\text{ndr},1}^{*} & f_{\text{ndr},2}^{*} &\cdots & f_{\text{ndr},s}^{*}&\cdots\\
\end{array}\right]
\end{align}
Now consider a fully connected neural network with $L$ hidden layers. Each layer uses a ReLU activation function denoted as $\sigma(\cdot)=\max(\cdot,0)$, and the output layer uses a linear activation function. We select ReLU as the activation function because it can simultaneously guarantee satisfactory training accuracy and bring computational advantages in terms of the reformulation. The predicted nadir can be expressed as follows:
\begin{align}
&\bold{z}_{1}=\bold{x}_{s}\bold{W}_{1}+\bold{b}_{1}\label{eq_nn_layer_in}\\
&\bold{\hat{z}}_{m}=\bold{z}_{m-1}\bold{W}_{m}+\bold{b}_{m}\label{eq_nn_layer_hidden}\\
&\bold{z}_{m}=\max(\bold{\hat{z}}_{m},0)\label{eq_nn_layer_ReLU}\\
&f_{\text{ndr},s}=\bold{z}_{N_{\text{Y}}}\bold{W}_{N_{\text{Y}}+1}+\bold{b}_{N_{\text{Y}}+1}\label{eq_nn_layer_out}
\end{align}
where the matrix $\bold{W}_{i}$ and vector $\bold{b}_{i}$ for $i=1,\cdots,N_{\text{Y}}$ represent the set of weight and bias across all hidden layers, and $\bold{W}_{N_{\text{Y}}+1}$ and $\bold{b}_{N_{\text{Y}}+1}$ represent the set of weight and bias of the output layer. We minimize the total mean squared error between the predicted output and the labeled outputs of all samples using the following:
\begin{align}
\min_{\bold{W}_{m}, \bold{b}_{m}}\frac{1}{N_{\text{S}}}\sum_{k=1}^{N_{\text{S}}}(f_{\text{ndr},s}-f^{*}_{\text{ndr},s})^{2}
\end{align}

\subsection{MILP Encoding of Trained Neural Networks}\label{subsec_dnn_encoding}
A binary vector $\bold{a}_{m}$ represents the activation status of ReLU at the $m$th hidden layer, and $\bold{a}_{m[n]}$ represents the status of the $n$th neuron at this layer. Let $[\bold{\underline{h}}_{m[n]},\bold{\overline{h}}_{m[n]}]$ be an interval that is large enough to contain all possible values of $\bold{\hat{z}}_{m[n]}$ in Eq. (\ref{eq_nn_layer_hidden}), where $\bold{\underline{h}}_{m[n]}<0$ and $\bold{\overline{h}}_{m[n]}>0$. Then the relation in Eq. (\ref{eq_nn_layer_ReLU}) can be expressed as:
\begin{align}
&\bold{z}_{m[n]}\leq \bold{\hat{z}}_{m[n]} - \bold{\underline{h}}_{m[n]}(1-\bold{a}_{m[n]})\label{eq_ReLU_1}\\
&\bold{z}_{m[n]}\geq \bold{\hat{z}}_{m[n]} \label{eq_ReLU_2}\\
&\bold{z}_{m[n]}\leq \bold{\overline{h}}_{m[n]}\bold{a}_{m[n]} \label{eq_ReLU_3}\\
&\bold{z}_{m[n]}\geq 0 \label{eq_ReLU_4}\\
&\bold{a}_{m[n]}\in\{0,1\} \label{eq_ReLU_5}
\end{align}
When $\bold{\hat{z}}_{m[n]}$ is less than or equal to zero, constraints (\ref{eq_ReLU_1}) and (\ref{eq_ReLU_4}) will force $\bold{a}_{m[n]}$ to be zero. In this case, constraints (\ref{eq_ReLU_3}) and (\ref{eq_ReLU_4}) imply that $\bold{z}_{i[k]}=0$, so we have $\bold{\hat{z}}_{i[k]}\leq 0 \implies \bold{a}_{m[n]}=0 \implies \bold{\hat{z}}_{m[n]}=0$. When $\bold{\hat{z}}_{m[n]}$ is greater than zero, constraints (\ref{eq_ReLU_2}) and (\ref{eq_ReLU_3}) will force $\bold{a}_{m[n]}$ to be 1. In this case, constraints (\ref{eq_ReLU_2}) and (\ref{eq_ReLU_3}) imply that $\bold{z}_{m[n]}=\bold{\hat{z}}_{m[n]}$, so we have $\bold{\hat{z}}_{m[n]}> 0 \implies \bold{a}_{m[n]}=1 \implies \bold{z}_{m[n]}=\bold{\hat{z}}_{m[n]}$. Obviously, this formulation contains no approximation of the original model. In addition, this is the tightest possible formulation with respect to its LP relaxation if no future information about $\bold{\hat{z}}_{m[n]}$ is revealed \cite{Anderson2019}.

\section{Microgrid Scheduling with Secure Islanding Capability}\label{sec_problem}
Let the microgrid be denoted as a graph $\mathcal{G}=(\mathcal{N}_{\text{B}},\mathcal{N}_{\text{L}})$, where $\mathcal{N}_{\text{B}}$ denotes all buses (vertices) and $\mathcal{N}_{\text{L}}$ denotes all lines (edges). Let $\mathcal{L}(\cdot,k)$ denote the set of lines for which bus $k$ is the to-bus, and $\mathcal{L}(k,\cdot)$ denote the set of lines for which bus $k$ is the from-bus. Let $\mathcal{D}(k)$, $\mathcal{W}(k)$ and $\mathcal{P}(k)$ define the sets of DGs, WTGs and substations connected to bus $k$, respectively. Let $\mu(l)$ and $\nu(l)$ map from the index of line $l$ to the index of its from-bus and to-bus, respectively. The nature of radiality guarantees that $\mu(l)$ and $\nu(l)$ are one-to-one mappings. The scheduling problem is formulated as follows. Bus 1 is assumed to be connected to the main grid.

Here a two-segment simplified cost, consisting of the fixed and marginal costs, is employed. 
The scheduling objective is to minimize the total operational cost, expressed as follows:
\begin{align}
\quad {\text{min}} \quad & \sum_{t=1}^{N_\text{T}}\sum_{i=1}^{N_{\text{D}}}\left[ \lambda^{\text{M}}_{i}p_{i,t}^{\text{D}}+\lambda^{\text{F}}_{i}u^{\text{D}}_{i,t}\right] \label{eq_obj_op}\\
&+ \sum_{t=2}^{N_\text{T}-1}\sum_{i=1}^{N_{\text{D}}}\lambda^{\text{S}}_{i}w^{\text{D}}_{i,t}\label{eq_obj_sp}\\
&+ \sum_{t=1}^{N_\text{T}}\rho^{\text{PCC}}_{t}P^{\text{PCC}}_{t}\label{eq_obj_buy}
\end{align}
Terms (\ref{eq_obj_op}) and (\ref{eq_obj_sp}) represent the fuel and start-up costs of diesel generators, respectively. Term (\ref{eq_obj_buy}) denotes the purchasing cost of energy that the microgrid operator pays to the main grid. It is worth noting that reformulation of the start-up cost has already been carried out in (\ref{eq_obj_sp}), where $w^{\text{D}}_{i,t}$ is the slack binary variable. In addition, $w^{\text{D}}_{i,t}$ and $u^{\text{D}}_{i,t}$ are subjected to the following constraints:
\begin{equation}
\begin{aligned}
&w^{\text{D}}_{i,t}\geq 0,w^{\text{D}}_{i,t}\geq u^{\text{D}}_{i,t}-u^{\text{D}}_{i,t-1}\quad\forall i,\forall t
\end{aligned}
\end{equation}

The power output of DSG $i$ equals the sum of the incremental output and minimum output, which is expressed in constraint (\ref{eq_DSG_1}). The sum of output and reserve should equal the dispatch upper bound, shown in constraint (\ref{eq_DSG_2}). The bound constraints of output and reserve are the formulations in (\ref{eq_DSG_3}) and (\ref{eq_DSG_4}), respectively. The binary status indicator is multiplied accordingly to ensure zero dispatchability when the unit is off.
\begin{align}
& P_{i,t}^{\text{D}}= p_{i,t}^{\text{D}} + \underline{P}_{i}^{\text{D}}u^{\text{D}}_{i,t}\qquad \forall i,\forall t\label{eq_DSG_1}\\
& p_{i,t}^{\text{D}}+R_{i,t}^{\text{D}}= (\overline{P}_{i}^{\text{D}}-\underline{P}_{i}^{\text{D}})u^{\text{D}}_{i,t}\qquad \forall i,\forall t\label{eq_DSG_2}\\
& 0\leq p_{i,t}^{\text{D}}\leq (\overline{P}_{i}^{\text{D}}-\underline{P}_{i}^{\text{D}})u^{\text{D}}_{i,t}\qquad \forall i,\forall t\label{eq_DSG_3}\\
& 0\leq R_{i,t}^{\text{D}}\leq (\overline{P}_{i}^{\text{D}}-\underline{P}_{i}^{\text{D}})u^{\text{D}}_{i,t}\qquad \forall i,\forall t\label{eq_DSG_4}
\end{align}

Following the convention in \cite{Wang2015} and \cite{Arif2018}, linearized distflow equations are employed to represent power flows in the network and are described as follows:
\begin{align}
& \begin{aligned}
\sum_{\forall l\in \mathcal{L}(\cdot,k)}P_{l,t} + \sum_{\forall i\in \mathcal{D}(k)}P^{\text{D}}_{i,t} &+ \sum_{\forall h\in \mathcal{P}(k)}P^{\text{PCC}}_{h,t}  + \sum_{\forall j\in \mathcal{W}(k)}P^{\text{W}}_{j,t}u^{\text{W}}_{j,t}\\
& = \sum_{\forall l\in \mathcal{L}(k,\cdot)}P_{l,t} + P^{\text{L}}_{k,t}\quad\forall k,\forall t
\end{aligned}\label{eq_dist_1}\\
& \sum_{\forall l\in \mathcal{L}(\cdot,k)}Q_{l,t} + \sum_{\forall i\in \mathcal{D}(k)}Q^{\text{D}}_{i,t} = \sum_{\forall l\in \mathcal{L}(k,\cdot)}Q_{l,t} + Q^{\text{L}}_{k,t}\quad\forall k,\forall t\label{eq_dist_2}\\
& V_{\nu(l),t}-V_{\mu(l),t} + \dfrac{R_{l}P_{l,t}+X_{l}Q_{l,t}}{V_{1}}=0 \quad\forall l,\forall t\label{eq_dist_3}\\
& 1-\epsilon\leq V_{k,t} \leq 1+\epsilon\quad\forall k,\forall t\label{eq_dist_4}\\
& \underline{P}_{l}\leq P_{l,t}\leq \overline{P}_{l}\quad\forall l,\forall t\label{eq_dist_5}\\
& \underline{Q}_{l}\leq Q_{l,t}\leq \overline{Q}_{l}\quad\forall l,\forall t\label{eq_dist_6}
\end{align}

Steady-state islanding capability concerns the power balance and frequency stability after islanding. To guarantee this capability, the total up- and down-spinning reserves in absolute value should be greater than the power at PCC:
\begin{align}
\label{eq_con_reserve}
\begin{aligned}
& \sum_{\forall h\in \mathcal{N}_{\text{P}}}P^{\text{PCC}}_{h,t} \leq \sum_{\forall i\in \mathcal{N}_{\text{D}}}R^{\text{D}}_{i,t}\quad\forall t\\
& \sum_{\forall h\in \mathcal{N}_{\text{P}}}P^{\text{PCC}}_{h,t} \geq -\sum_{\forall i\in \mathcal{N}_{\text{D}}}p^{\text{D}}_{i,t}\quad\forall t
\end{aligned}
\end{align}

For the safe operation of WTGs, the inertia emulation functions can be activated only when the WTG is at a certain percentage of fully loaded condition. This percentage is to guarantee adequate available energy stored in WTGs to perform the inertia emulation control. Detailed calculations can be found in \cite{Wang2018a} and \cite{Wang2019} and are omitted here. This constraint can be formulated as follows:
\begin{align}
& e^{\text{W}}_{j,t} \geq (P^{\text{W}}_{j,t} - \alpha\overline{P}^{\text{W}}_{j})/M\qquad \forall i,\forall t\label{eq_IE_1}\\
& e^{\text{W}}_{j,t} < (P^{\text{W}}_{j,t} - \alpha\overline{P}^{\text{W}}_{j})/M + 1\qquad \forall i,\forall t\label{eq_IE_2}\\
& u^{\text{IE}}_{j,t} \leq e^{\text{W}}_{j,t}\qquad \qquad \qquad \forall i,\forall t\label{eq_IE_3}\\
& u^{\text{IE}}_{j,t} \leq u^{\text{W}}_{j,t},u^{\text{IE}}_{j,t} \leq u^{\text{W}}_{j,t}\qquad \forall i,\forall t\label{eq_IE_4}
\end{align}
where a slack binary variable $e^{\text{W}}_{j,t}$ is introduced to represent the availability of the inertia emulation function, and $M$ should hold the condition $|P^{\text{W}}_{j,t} - \alpha\overline{P}^{\text{W}}_{j}|<M$. When a WTG reaches the permissible condition, we have $P^{\text{W}}_{j,t} - \alpha\overline{P}^{\text{W}}_{j}\geq 0$. Under these circumstances, constraints (\ref{eq_IE_1}) and (\ref{eq_IE_2}) force the binary variable $e^{\text{W}}_{j,t}$ to be 1. However, if $P^{\text{W}}_{j,t} - \alpha\overline{P}^{\text{W}}_{j}<0$, $e^{\text{W}}_{j,t}$ will be forced to be zero.
Constraint (\ref{eq_IE_3}) ensures that the inertia emulation can be activated only when the WTG operates at the permissible range. Constraint (\ref{eq_IE_4}) ensures that the inertia emulation can be activated only when the WTG is on.

The predicted frequency nadir should be limited:
\begin{align}
& f_{\text{ndr},t}\geq f^{\text{UFLS}}_{\text{ndr}}\quad\forall t\label{eq_con_dyn1}
\end{align}
The predicted frequency nadir depends on the system operating condition defined in the following vector:
\begin{align}
\bold{x}_{t}=\left[u^{\text{D}}_{1,t},\cdots,u^{\text{D}}_{N_{\text{D}},t},\sum_{j}u^{\text{IE}}_{j,t},P^{\text{PCC}}_{t}\right]\label{eq_con_dyn2}
\end{align}
The nadir point is then obtained using the MILP formulation of the neural network in Section \ref{subsec_dnn_encoding}:
\begin{equation}
\begin{aligned}
&\bold{z}_{1,t}=\bold{x}_{t}\bold{W}_{1}+\bold{b}_{1}  \quad\forall t\\
&\bold{\hat{z}}_{m,t}=\bold{z}_{m-1,t}\bold{W}_{m}+\bold{b}_{m}  \quad\forall m,\forall n,\forall t\\
&\bold{z}_{m[n],t}\leq \bold{\hat{z}}_{m[n],t} - \bold{\underline{h}}_{m[n]}(1-\bold{a}_{m[n],t})  \quad\forall m,\forall n,\forall t\\
&\bold{z}_{m[n],t}\geq \bold{\hat{z}}_{m[n],t}  \quad\forall m,\forall n,\forall t\\
&\bold{z}_{m[n],t}\leq \bold{\overline{h}}_{m[n]}\bold{a}_{m[n],t} \quad\forall m,\forall n,\forall t\\
&\bold{z}_{m[n],t}\geq 0 \quad\forall m,\forall n,\forall t\\
&f_{\text{ndr},t}=\bold{z}_{N_{\text{Y}},t}\bold{W}_{N_{\text{Y}}+1}+\bold{b}_{N_{\text{Y}}+1} \quad\forall t\\
&\bold{a}_{m,t}\in\{0,1\}
\end{aligned}\label{eq_con_dyn3}
\end{equation}
Note that the notations of variables are the same, except that we need to define the output $\bold{z}_{m}$ and ReLU activation $\bold{a}_{m}$ of all layers for each period $t$.

\section{Case Study}\label{sec_case}
\begin{figure}[h]
	\centering
	\includegraphics[scale=0.25]{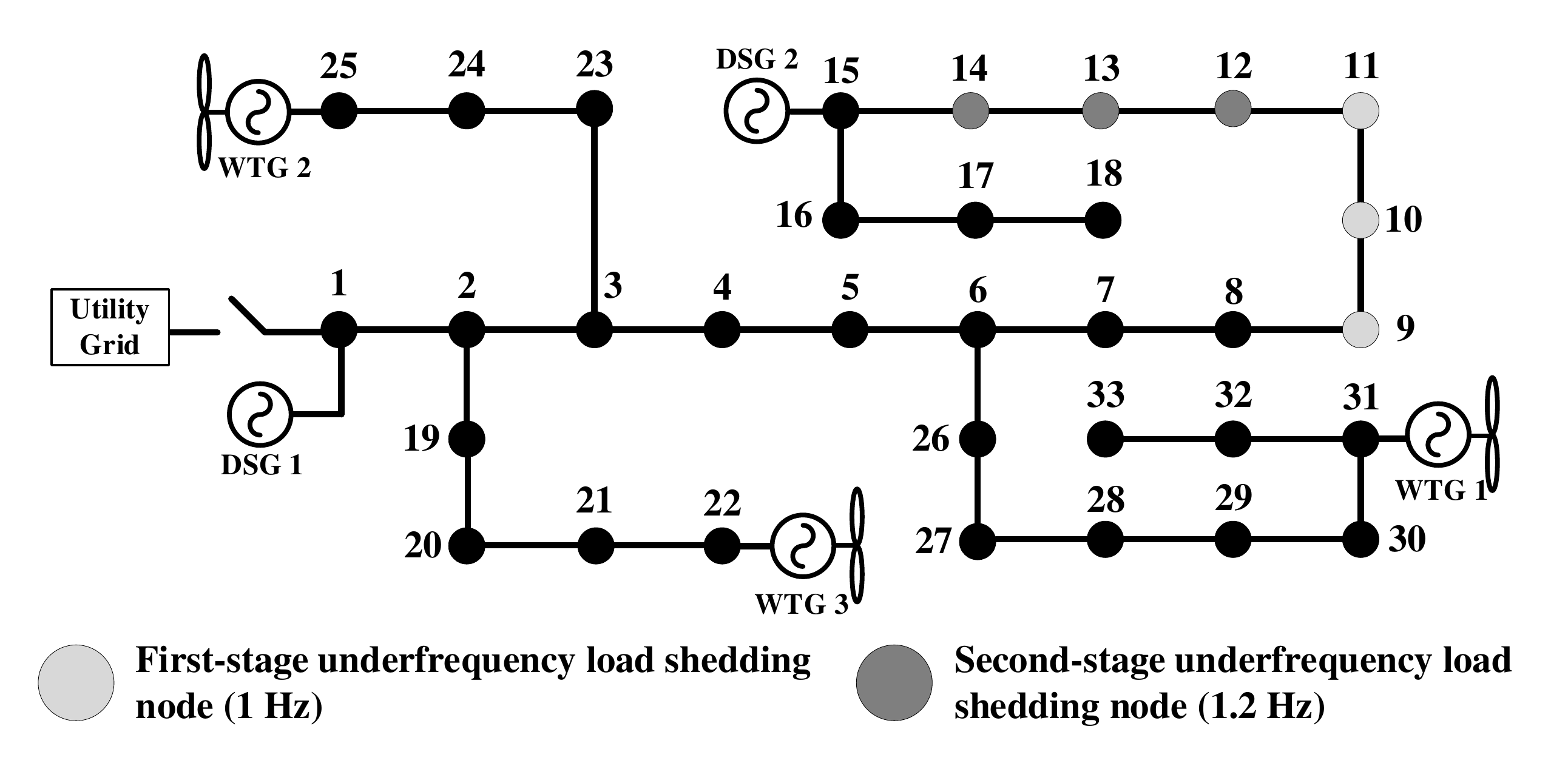}
	\caption{The modified 33-node system with DSGs, WTGs and a two-stage underfrequency load shedding strategy.}
	\label{fig_33Feeder}
\end{figure}

The modified 33-node system in \cite{Wang2019b}, which has been widely adopted for microgrid studies \cite{Wang2015}\cite{Gholami2019}, will be employed in this paper. It is a radial 12.66 kV distribution network, shown in Fig. \ref{fig_33Feeder}. The forecast total load and wind power over the 24-hour period are given in MW in Table \ref{tab_forecast_data}. The total load is distributed to each bus according to the load profile in \cite{Stentz1988}. The DSGs are connected to buses 1 and 15, while WTGs are connected to buses 22, 25 and 31. Detailed network data can be found in \cite{Stentz1988}.  Day-ahead market prices (in ct/kWh) are adopted from \cite{Liu2017} and listed in Table \ref{tab_forecast_data}.
The key parameters of DSGs are given in Table \ref{tab_diesel_data}. For both DSGs, $\tau_{d}=0.1$, $\tau_{sm}=0.5$ and $R_{D}=0.05$. The three  WTGs are identical, with 400 kW rated power. The inertia emulation gain is set as $K_{\text{ie}}=0.1$ with a dead-band $[59.85,65]$. 
\begin{table}[H]
	\caption{Forecast Data of Load, Wind Power and Electricity Price}
	\centering
	\begin{tabular}{lclclclclclclclcl}
		\toprule 
		$\bold{Period}$ & Load & Wind & Price  & $\bold{Period}$ & Load & Wind & Price\\
		\midrule
		$\mathbf{1}$ & 2.210 & 12.3 & 8.65 & $\mathbf{13}$ & 3.367 & 9.1 & 26.82 \\
		$\mathbf{2}$ & 2.197 & 11.8 & 8.11 & $\mathbf{14}$ & 3.315 & 10.2 & 27.35  \\
		$\mathbf{3}$ & 2.249 & 12.2 & 8.25 & $\mathbf{15}$ & 3.406 & 11.3 & 13.81 \\ 
		$\mathbf{4}$ & 2.210 & 10.4 & 8.10  & $\mathbf{16}$ & 3.445 & 12.0 & 17.31 \\
		$\mathbf{5}$ & 2.275 & 10.5 & 8.14 & $\mathbf{17}$ & 3.315 & 11.7 & 16.42  \\
		$\mathbf{6}$ & 2.405 & 10.3 & 8.13 & $\mathbf{18}$ & 3.289 & 11.5 & 9.83 \\
		$\mathbf{7}$ & 2.600 & 9.3 & 8.34  & $\mathbf{19}$ & 3.250 & 9.4 & 8.63  \\
		$\mathbf{8}$ & 3.159 & 9.7 & 9.35 & $\mathbf{20}$ & 3.315 & 9.6 & 8.87 \\
		$\mathbf{9}$ & 3.302 & 8.5 & 12.00  & $\mathbf{21}$ & 3.380 & 10.1 & 8.35  \\
		$\mathbf{10}$ & 3.380 &9.2 & 9.19 & $\mathbf{22}$ & 3.224 & 11.3 & 16.44 \\
		$\mathbf{11}$ & 3.471 & 8.7 & 12.30  & $\mathbf{23}$ & 2.960 & 12.2 & 16.19  \\
		$\mathbf{12}$ & 3.367 & 7.9 & 20.70 & $\mathbf{24}$ & 2.392 & 11.7 & 8.87 \\
		\bottomrule
	\end{tabular}
	\label{tab_forecast_data}
\end{table}

As we mentioned in Section \ref{sec_intro}, the motivation for considering frequency trajectory constraints is to avoid underfrequency load shedding so the customer supply is not interrupted. Here, we set our frequency excursion limit based on the WECC standard \cite{WECC_load_shedding}. The detailed WECC underfrequency load shedding logic is to launch the first stage of the plan when the frequency excursion is larger than 0.9 Hz for 14 cycles, so we use 1 Hz as the frequency trajectory limit to approximate this temporal logic.

For comparison purposes, we also implement the two-stage underfrequency load shedding approach shown in Fig. \ref{fig_33Feeder}. In the first stage, loads at Bus 9-11 will be shed when the frequency deviation is larger than 1 Hz. The second stage will shed the loads at Bus 12-14 when the frequency deviation is larger than 1.2 Hz.

The optimization is formulated using Pyomo \cite{hart2017pyomo} and solved using IBM ILOG CPLEX 12.8. The dynamic simulation is performed using the TPN model in the Simulink environment. The deep learning model is built using TensorFlow $r1.14$ \cite{tensorflow2015-whitepaper}.

\begin{table}[H]
	\caption{Diesel Generator Data}
	\centering
	\begin{tabular}{lclclclclclclcl}
		\toprule 
		\# & Base & $H_{D}$  & $[\underline{P}^{D},\overline{P}^{D}]$ & $\lambda^{\text{M}}$& $\lambda^{\text{F}}$& $\lambda^{\text{S}}$\\
		\midrule
		1 & 1 [MW] & 4 & [0.2,1] & 3.32 & 0.026 & 3\\
		2 & 2 [MW] & 3 & [0.4,2]  & 2.55 & 0.033 & 1\\
		\bottomrule
	\end{tabular}
	\label{tab_diesel_data}
\end{table}

\subsection{Frequency Nadir Predictor}\label{sec_sub_dnn_pred}
The model shown in Fig. \ref{fig_PSPB} is used to generate the training data. WTGs are operated at rated condition. The PCC power is generated randomly from the uniform distribution in the interval $[-2, 2]$. In each sampled PCC power, we will consider different combinations of DSG status and number of activated inertia emulation functions, which are denoted by $u^{\text{D}}_{i,s}$ and $\sum_{j}u^{\text{IE}}_{j,s}$, respectively. Since at least one DSG will stay committed to provide frequency and voltage regulation during islanding, there are 12 scenarios for each sampled PCC power. We obtained a total of 4500 observations (samples), $80\%$ of which are used to train the neural network (the rest are for testing purposes). Thanks to the PSPB model, it only takes around 15 to 50 seconds to generate one sample, and in total 18.75 hours to generate all samples. The TPN model spends around 600 seconds for one scenario, and will require 750 hours.

\begin{figure}[h]
	\centering
	\includegraphics[scale=0.27]{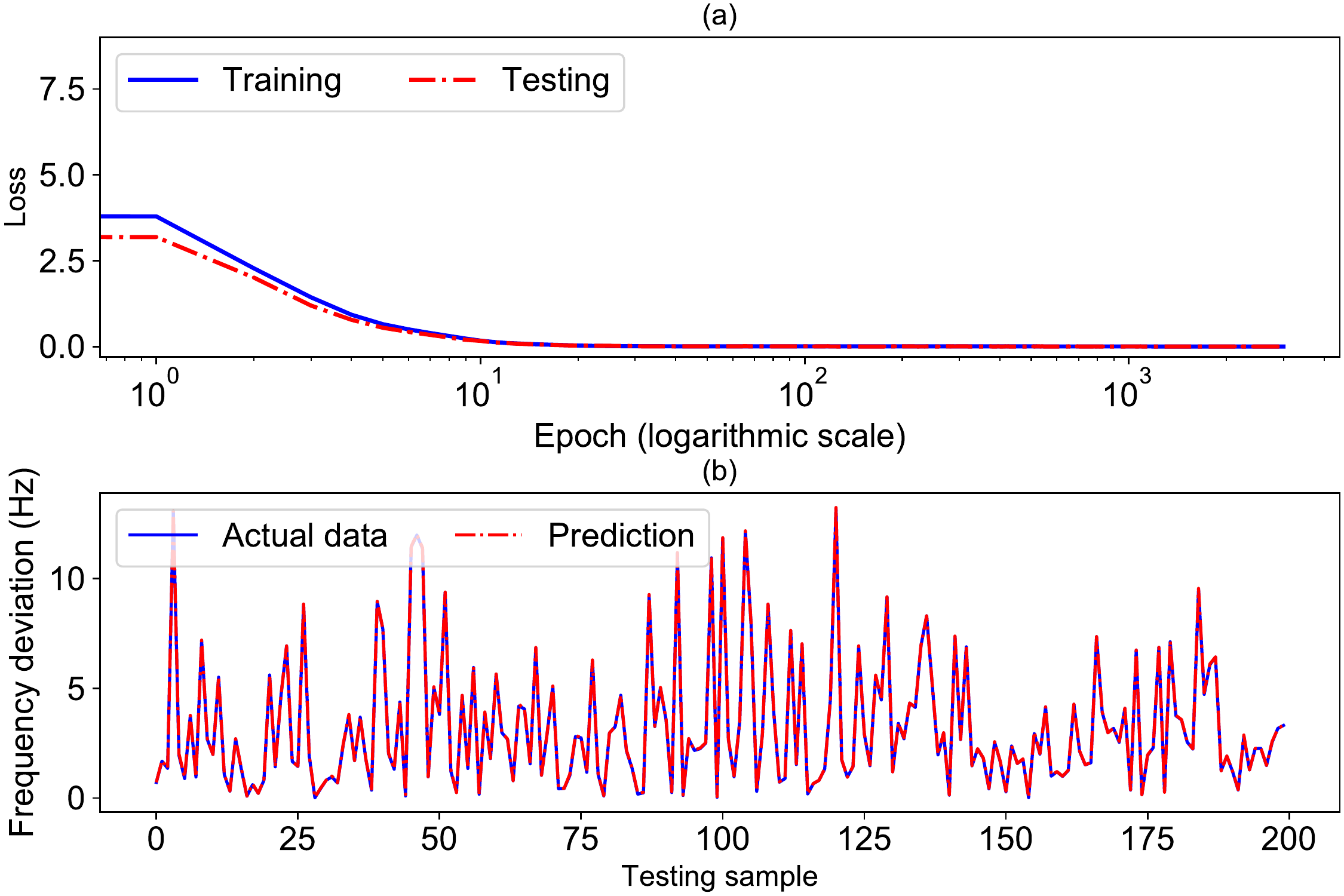}
	\caption{(a) Loss during neural network training. (b) Testing result.}
	\label{fig_sample}
\end{figure}

The neural network has one hidden layer with 40 neutrons. The training and testing results are plotted in Fig. \ref{fig_sample} (a) and (b), respectively. The number of epochs is plotted in logarithmic scale. The total mean squared error between the predicted and labeled outputs of all samples converges to zero. Fig. \ref{fig_sample} (b) shows the precise prediction of the trained model using testing data.

\subsection{Scheduling with Islanding Capability}\label{sec_sub_schedule}
\begin{figure}[h]
	\centering
	\includegraphics[scale=0.27]{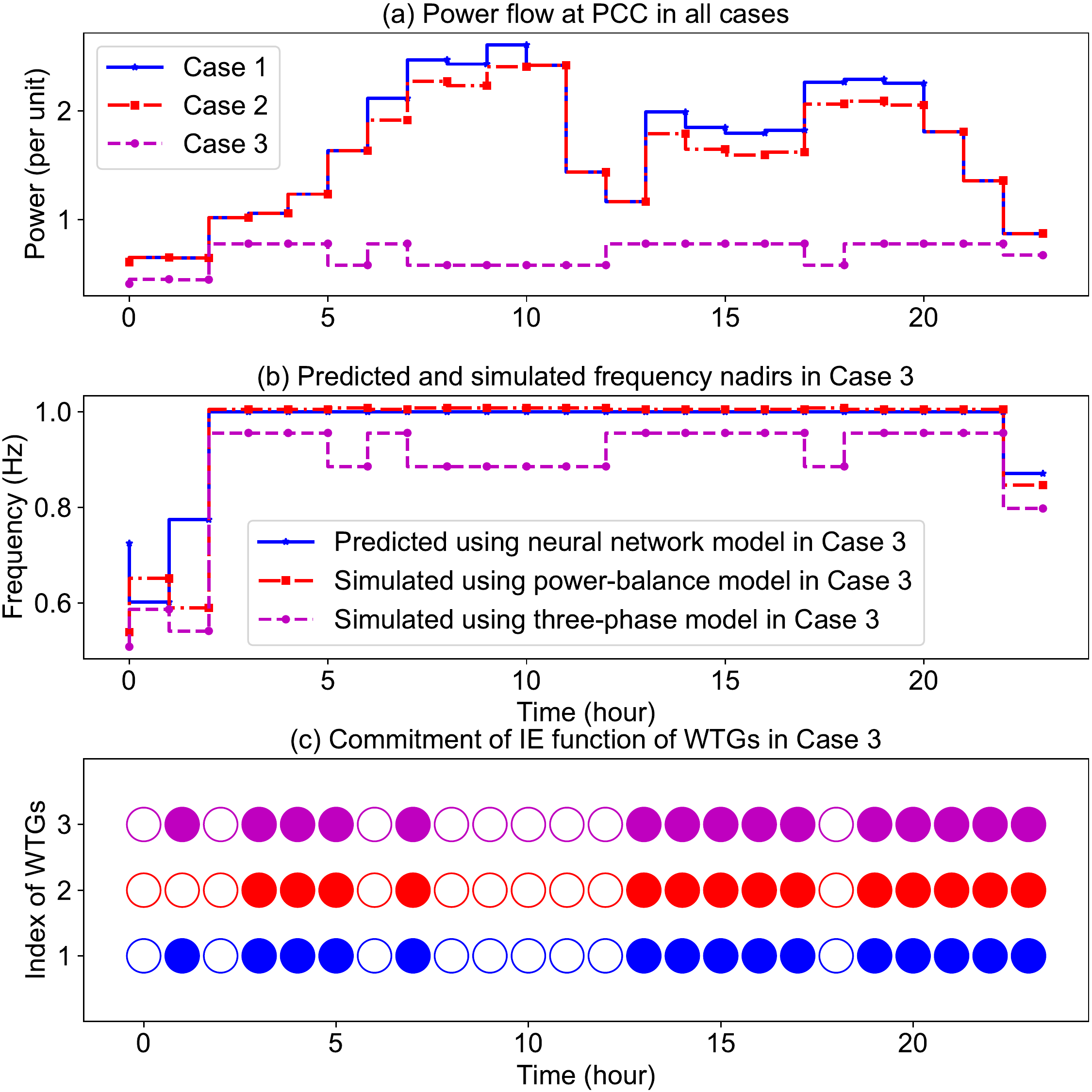}
	\caption{(a) Power at PCC in different cases. (b) Predicted frequency nadir compared with two simulated by PSPB and TPH models. (c) Scheduling of WTG inertia emulation functions,  where filled circles denote activate and unfilled ones denote inactivate.}
	\label{fig_scheduling_1}
\end{figure}

\begin{figure}[h]
	\centering
	\includegraphics[scale=0.27]{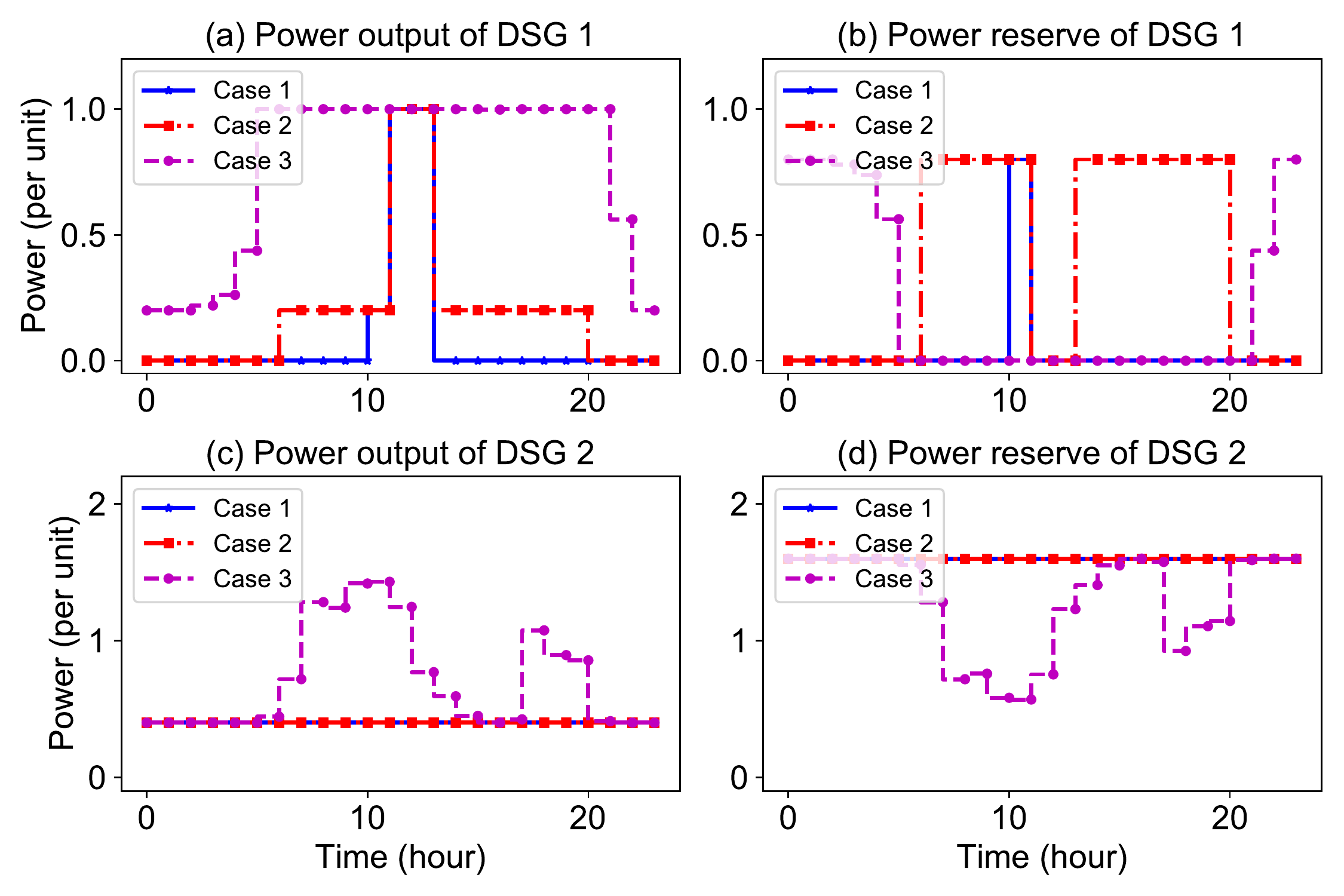}
	\caption{Power output and reserve of DSGs under different cases.}
	\label{fig_scheduling_2}
\end{figure}

\begin{figure}[h]
	\centering
	\includegraphics[scale=0.27]{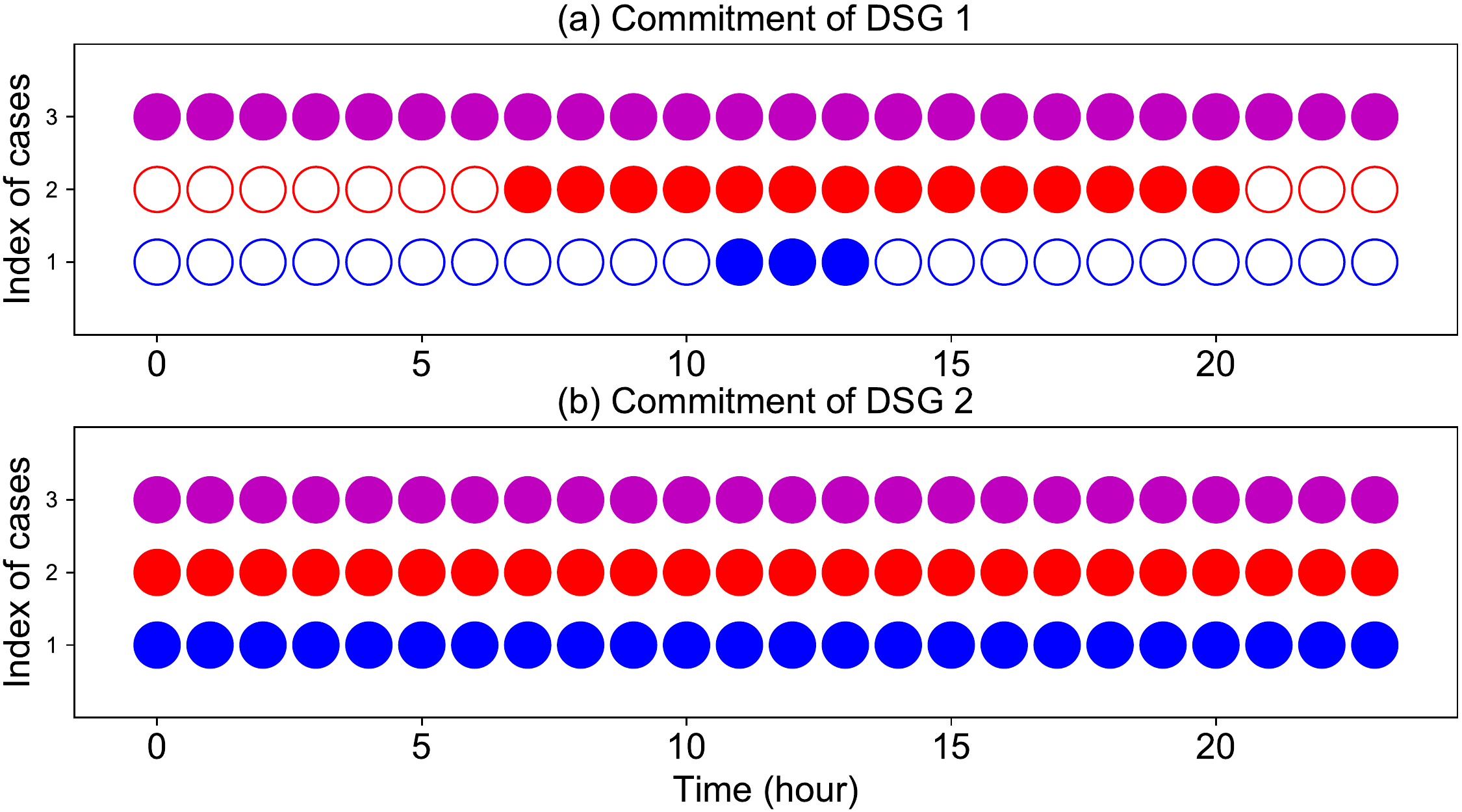}
	\caption{Commitment of DSGs under different cases, where filled circles denote on and unfilled ones denote off.}
	\label{fig_scheduling_3}
\end{figure}

We will consider three different cases as follows: 
 \begin{enumerate}
 	\item Scheduling without islanding constraints with the problem formulation using Eqs. (\ref{eq_obj_op}) - (\ref{eq_DSG_4})
 	\item Scheduling with static islanding constraints with the problem formulation using Eqs. (\ref{eq_obj_op}) - (\ref{eq_con_reserve})
 	\item Scheduling with dynamic islanding constraints with the problem formulation using Eqs. (\ref{eq_obj_op}) - (\ref{eq_con_dyn3})
 \end{enumerate}
The power at PCC for all three cases is plotted in Fig. \ref{fig_scheduling_1} (a). In Case 3, the predicted frequency nadirs are compared with simulated nadirs from PSPB and TPN models as shown in Fig. \ref{fig_scheduling_1} (b).  The scheduling of inertia emulation functions is shown in Fig. \ref{fig_scheduling_1} (c). The power and reserve scheduling results are shown in Fig. \ref{fig_scheduling_2}. The commitment results of DGSs are plotted in Fig. \ref{fig_scheduling_3}. The frequency nadirs obtained by three different methods show high consistency, indicating that (1) the prediction performance of the trained neural network is satisfactory even when complex factors have been considered, and (2) the PSPB model can precisely exhibit the frequency behaviors of the system.
The PCC power in Fig. \ref{fig_scheduling_1} (a) indicates that purchasing energy from the utility grid is preferred in Cases 1 and 2. The DSGs are required to complement the load during Periods 12-14 due to the high electricity price, as shown in Fig. \ref{fig_scheduling_2}. Compared with Case 1, Case 2 has a relatively smaller PCC power and a longer commitment period for DSG 1, during which DSG 1 is providing the spinning reserve.  The frequency nadir constraint, however, confines the PCC power to a much smaller value, so that the smaller disturbance will be imposed on the frequency control system once an islanding event takes place. Observing Fig. \ref{fig_scheduling_1} yields the conclusion that the frequency nadir constraint is binding from Period 4 to Period 23. All inertia emulation functions are scheduled upon availability. The largest PCC power without inertia emulation support is 0.59 MW, becoming 0.79 MW when all inertia emulation functions are online. This will reduce the output of DSG 2 and make the system purchase cheaper energy from the main grid. In Periods 1, 2, 3 and 24, the binding constraint is the minimal output constraint of DSGs. To maintain adequate inertial response, DSG 1 is committed for all periods shown in Fig. \ref{fig_scheduling_3} and must operate at  minimal power even when the load consumption is small, resulting in a small PCC power and therefore a smaller frequency nadir. 

\begin{figure}[h]
	\centering
	\includegraphics[scale=0.3]{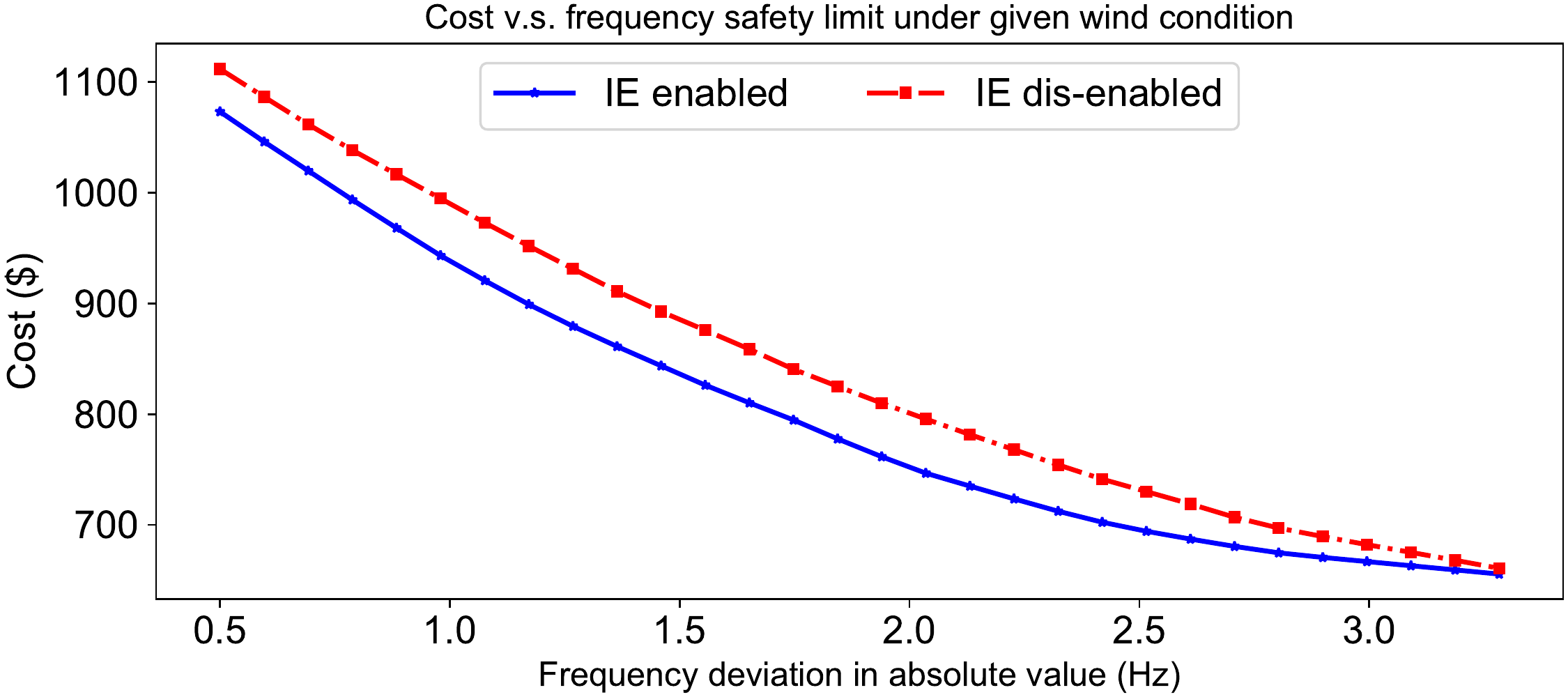}
	\caption{Operation cost with respect to different frequency nadir requirements.}
	\label{fig_scheduling_cost}
\end{figure}

The total operational cost over 24 periods for different frequency constraint specifications under a given wind condition is plotted in Fig. \ref{fig_scheduling_cost}. For illustration purposes, the frequency constraint is extended to large (unsafe) values. As shown, when the nadir is larger than 4.2 Hz, this constraint produces no extra cost. Costs when the inertia emulation functions are dis-enabled are also compared. In the nadir range from 0.5 to 3 Hz, the dis-enablement of inertia emulation functions results in a $38\%$ increase in the operational cost. 

\subsection{TPN Model-Based Simulation Verification}\label{sec_sub_verification}
\begin{figure}[h]
	\centering
	\includegraphics[scale=0.35]{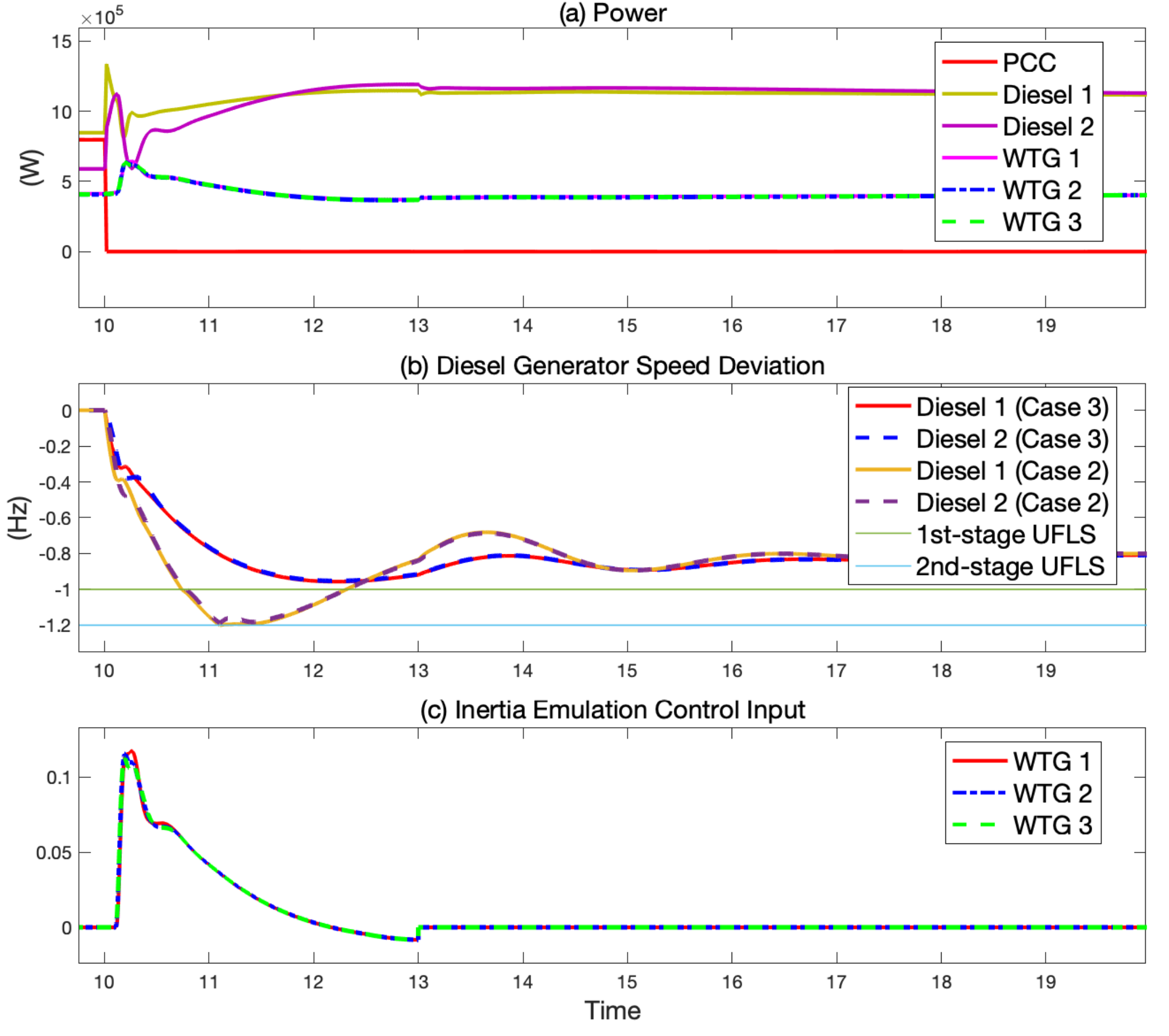}
	\caption{System dynamic responses to the islanding event in Period 8. (a) Power dynamics. (b) DSG speed deviations in the scheduling conditions of Case 2 and Case 3. (c) Control signal of inertia emulation.}
	\label{fig_System_Responses}
\end{figure}

To show that the scheduling commands of Case 3 can lead to successful and secure islanding, we assume that a severe fault occurs in the main grid during Period 8, and event-triggered islanding is executed under the operating conditions given in Case 3. The system dynamic responses are shown in Fig. \ref{fig_System_Responses}, including power dynamics of different sources, DSG speed deviations, and the control signal for inertia emulation. The DSG speed trajectories are secure with no load shedding plans being triggered, as the trajectory constraints effectively limit the PCC power, as shown in Fig. \ref{fig_scheduling_1} (a). For comparison, we use the static islanding constrained formulations (Case 2) to schedule the microgrid and assume the same islanding event during Period 8. The DSG speed deviations in this scenario are plotted in Fig. \ref{fig_System_Responses} (b). Since the static islanding constraints admit a larger PCC power, the DSG speed deviations first exceed the first-stage load shedding threshold, which reduces their rate of change, and then pass the second-stage load shedding threshold, which finally arrests the decline.

\begin{table}[H]
	\caption{Scheduling Periods with Frequency Nadirs by Different Methods}
	\centering
	\begin{tabular}{lclclclclclclclcl}
		\toprule 
		$\bold{Period}$ & Predicted & Linear (Error) & Nonlinear (Error) \\
		\midrule
		1 & 									 0.532 & 0.539 (1.33\%) & 0.509 (4.55\%)   \\
		2 & 									 0.644 & 0.652 (1.13\%) & 0.587 (9.78\%)    \\
		3 & 									 0.583 & 0.590 (1.29\%) & 0.542 (7.52\%)   \\ 
		4--6,8,14-18,20--23 & 		 1.000 & 1.005 (0.49\%) & 0.956 (4.60\%)    \\
		7,9--13,18 & 					  1.000 & 1.008 (0.84\%)  & 0.886 (12.86\%)   \\
		24 & 									 0.843 & 0.847 (0.51\%) & 0.798 (5.58\%)  \\
		\bottomrule
	\end{tabular}
	\label{tab_result}
\end{table}
\begin{figure}[h]
	\centering
	\includegraphics[scale=0.35]{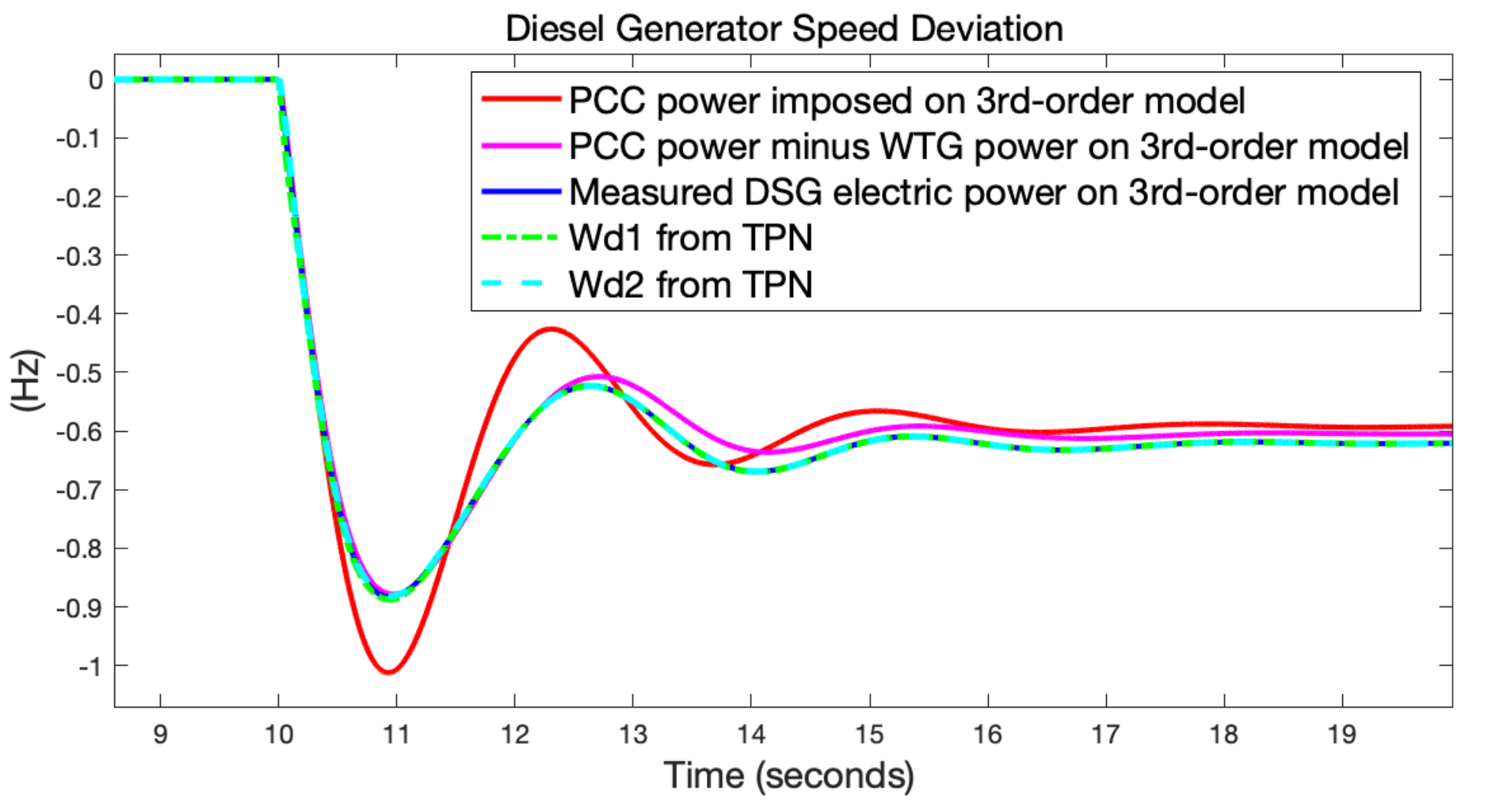}
	\caption{Frequency responses to feeding different combinations of nonlinear simulated power to the response model compared with nonlinear simulation results.}
	\label{fig_Frequency}
\end{figure}

In Case 3, all periods can be categorized into six scenarios, based on the PCC power. Frequency nadirs of these scenarios obtained by neural network, linear model and nonlinear model simulations are listed in Table \ref{tab_result}. As shown, the percentage error of the linear model is, on average, $0.92\%$. This indicates the correctness and effectiveness of the machine learning technique presented. The percentage error of the nonlinear model, however, varies with different cases. For those cases in which all inertia emulation functions are activated, the average error is $4.90\%$, which is satisfactory considering the complexity of the nonlinear model. For those cases in which all inertia emulation functions are deactivated, the averaged error is $12.86\%$. We assume that this is because the DFIG-based WTG is not fully decoupled and admits a weak inertial response \cite{Zhang2018a}\cite{Mullane2005}, which is not captured by the linear model.

To validate this argument, we simulate an islanding in Period 7 and impose simulated power on Eq. (\ref{eq_DSG}) in different combinations. First, the simulated PCC power variation $\Delta \widetilde{P}^{\text{PCC}}$ is imposed on Eq. (\ref{eq_DSG}) as:
\begin{equation}
\begin{aligned}
& \Delta P_{e} = \Delta \widetilde{P}^{\text{PCC}}
\end{aligned}\label{eq_power_verify_1}
\end{equation}
where the tilde symbol denotes that the simulated data is from the TPN model. Second, the variational power of WTGs is subtracted:
\begin{equation}
\begin{aligned}
& \Delta P_{e} = \Delta \widetilde{P}^{\text{PCC}} - \sum_{j\in\mathcal{N}_{\text{w}}}\Delta \widetilde{P}_{g,j}
\end{aligned}\label{eq_power_verify_2}
\end{equation}
Third, the total simulated electric power variation of DSGs is imposed on Eq. (\ref{eq_DSG}):
\begin{equation}
\begin{aligned}
& \Delta P_{e} = \Delta \widetilde{P}_{e}
\end{aligned}\label{eq_power_verify_3}
\end{equation}
The three frequency trajectories are shown in Fig. \ref{fig_Frequency}, together with the ones from the TPN simulation. All other cases show high consistency, except for the first case, where the weak inertial responses from WTGs are not taken into account. Fortunately, this simplification will lead to conservative scheduling and pose no security concern.

This comparative study also verifies the accuracy of Eq. (\ref{eq_DSG}). As shown in Fig. \ref{fig_Frequency}, when the same electric power variation is imposed on Eq. (\ref{eq_DSG}) and full-order DSG, their trajectories coincide with each other. This implies that the response model in (\ref{eq_DSG}) can sufficiently represent the frequency response of a DSG.

\section{Conclusions}
This paper presents a microgrid scheduling problem with frequency-constrained islanding capability. The nonlinear function between system operating condition and frequency nadir is approximated using a neural network. Due to its strong representation power, realistic factors such as grid-interactive converters, dead-band, saturation, and low-pass filters can be considered. More significantly, the trained neural network admits an exact mixed-integer formulation. To efficiently generate training data from simulations, a PSPB model is derived from the original TPN model. Simulation time can be reduced by $97.5\%$ with sufficient accuracy in the representation of frequency response. The resulting MIP is integrated into the scheduling problem to encode the frequency constraint. The proposed method is validated on the modified 33-node system using a detailed three-phase model in Simulink. The dispatch and control commands ensure both islanding success and adequate frequency response. In addition, inertia emulation functions are able to reduce the operation cost. It is worth noting that the deep learning based constraint encoding technique presented can be employed for any dynamic-constrained optimization problem. 
Future work will seek advanced optimization approaches that can best utilize state-of-the-art grid-forming converter technology \cite{Ma2017}. In addition, different types of load models will be considered, particularly controllable loads such as variable speed motor drives. When the information about the load model is incomplete, the reinforcement learning framework could be employed \cite{Zhang2020}.

\bibliography{IEEEabrv_zyc,library,Ref_DCS}
\bibliographystyle{IEEEtran}

\end{document}